\documentclass[aip,rsi,amsmath,amssymb,reprint]{revtex4-1}

\usepackage{graphicx}
\usepackage{dcolumn}
\usepackage{bm}
\usepackage[mathlines]{lineno}
\usepackage[utf8]{inputenc}
\usepackage[T1]{fontenc}
\usepackage{mathptmx}
\usepackage{etoolbox}
\usepackage{xcolor}
\usepackage{textcomp}
\usepackage[version=4]{mhchem}

\usepackage{float}
\floatstyle{plaintop}
\restylefloat{table}

\newcommand{\Rd}{\ensuremath{R_d}}
\newcommand{\Rv}{\ensuremath{R_{\text{v}}}}
\newcommand{\Cp}{\ensuremath{C_p}}

\newcommand{\RH}{\ensuremath{S_0}}
\newcommand{\po}{\ensuremath{p_0}}
\newcommand{\To}{\ensuremath{T_0}}
\newcommand{\pf}{\ensuremath{p_f}}
\newcommand{\Tf}{\ensuremath{T_f}}
\newcommand{\psat}{\ensuremath{p_{\text{sat}}}}
\newcommand{\Tsat}{\ensuremath{T_{\text{sat}}}}
\newcommand{\qsat}{\ensuremath{q_{\text{sat}}}}
\newcommand{\ysat}{\ensuremath{y_{\text{sat}}}}

\newcommand{\dbar}{\ensuremath{\bar{d}}}

\makeatletter
\def\@email#1#2{
 \endgroup
 \patchcmd{\titleblock@produce}
  {\frontmatter@RRAPformat}
  {\frontmatter@RRAPformat{\produce@RRAP{*#1\href{mailto:#2}{#2}}}\frontmatter@RRAPformat}
  {}{}
}
\makeatother
\begin{document}

\preprint{AIP/123-QED}
\title[Droplet Nucleation In a Rapid Expansion Aerosol Chamber]{Droplet Nucleation In a Rapid Expansion Aerosol Chamber}

\author{Martin A. Erinin}
\affiliation{Department of Mechanical \& Aerospace Engineering, Princeton University, New Jersey, USA}

\author{Cole R. Sagan}
\affiliation{Department of Chemistry, Princeton University, New Jersey, USA}

\author{Ilian Ahmed}
\affiliation{Department of Mechanical \& Aerospace Engineering, Princeton University, New Jersey, USA}
\affiliation{Département de Physique, École Normale Supérieure de Paris, Paris, France}

\author{Gwenore F. Pokrifka}
\affiliation{High Meadows Environmental Institute, Princeton University, New Jersey, USA}

\author{Nadir Jeevanjee}
\affiliation{Geophysical Fluid Dynamics Laboratory, NOAA, Princeton, New Jersey, USA}

\author{Marissa L. Weichman}
\affiliation{Department of Chemistry, Princeton University, New Jersey, USA}

\author{Luc Deike$^{*,}$}
\email{ldeike@princeton.edu}
\affiliation{Department of Mechanical \& Aerospace Engineering, Princeton University, New Jersey, USA}
\affiliation{High Meadows Environmental Institute, Princeton University, New Jersey, USA}

\date{\today}

\begin{abstract}
We present a new experimental facility to investigate the nucleation and growth of liquid droplets and ice particles under controlled conditions and characterize processes relevant to cloud microphysics: 
the rapid expansion aerosol chamber (REACh). REACh is an intermediate size chamber ($\sim0.14$~m$^3$) combining the principle of an expansion chamber with the ability to probe the influence of turbulent flows. Nucleation is achieved via a sudden pressure drop accompanied by a temperature drop, which can cause humid air to condense into a cloud of droplets under the appropriate thermodynamic conditions.
REACh features tight control and monitoring of the initial saturation ratio of water vapor, identity and concentration of seeding aerosol particles, temperature, pressure, and air flow mixing, together with high speed real time measurements of aerosol and droplet size and number. 
Here, we demonstrate that the minimum temperature reached during each expansion can be reasonably described by the thermodynamics of dry or moist  adiabats, for a wide range of initial relative humidity. 
The size and number of droplets formed, and the overall lifetime of the cloud, are characterized as a function of the aerosol concentration and initial water vapor saturation ratio.  
The total droplet concentration scales linearly with the seeding aerosol concentration, suggesting that all injected aerosol particles serve as condensation nuclei.  While the total number of droplets formed increases with aerosol concentration, the mean droplet size decreases with the concentration of seeding aerosols as a result of competition for the available water vapor. Theoretical considerations provide a quantitative prediction for the mean droplet size for a wide range of conditions.
The high repetition rate of experiments that we can perform in REACh will permit extensive characterization of aerosol-nucleation including nucleation onset, droplet and ice growth and the importance of turbulence fluctuations. We will leverage the facility's capabilities to explore a wide range of physical parameters encompassing regimes relevant to cloud microphysics .
\end{abstract}

\maketitle

\section{Introduction}\label{sec:intro}


One of the largest remaining uncertainties in climate projections involves the impact of aerosols on the optical properties of clouds\citep{calvin2023ipcc}. 
Aerosols are particles suspended in the atmosphere that can serve as cloud condensation nuclei (CCN) upon which water vapor can condense in the formation of cloud droplets .
Understanding the influence of different seeding aerosols on droplet nucleation, growth, and evaporation is essential for accurate predictions of cloud behavior and the resulting impact on the environment \citep{ghan2011droplet,seinfeld1998air,pruppacher1998microphysics}.

The relevant physical scales for warm clouds formed by liquid droplets span several orders of magnitude, ranging from nano-scale aerosol particles ($\mathcal{O}$(10~nm)), to micro-scale droplets ($\mathcal{O}$(10~$\mu$m)), to centimeter-scale turbulence fluctuations, 1-10m scale turbulent processes like cloud top entrainment and up to large scale convective patterns ($\mathcal{O}$(1-10~km)). 
The nucleation of cloud droplets and their subsequent growth dynamics lead to specific concentrations and size distributions of droplets, depending on the seeding aerosol number density, composition, as well as the surrounding thermodynamic conditions (e.g. water vapor, temperature, and pressure), which are themselves modulated by the surrounding turbulent flow \citep{bodenschatz2010can,prabhakaran2020role,yeom2023cloud}. These factors determine whether a cloud droplets will remain stable for extended periods of time or rain out, and also govern the optical interactions of the droplets with sunlight and terrestrial radiation \citep{pruppacher1998microphysics,poydenot2024gap,poydenot2024pathways}.

Beyond the challenges of droplet nucleation and properties of warm turbulent clouds, higher latitude low level clouds can be mixed-phase, composed of mixtures of crystalline ice and liquid water droplets\citep{burrows2022ice,storelvmo2017aerosol} adding a further layer of complexity to aerosol-nucleation microphysics. At higher altitudes, cirrus clouds and contrails are formed purely of ice particles. The conditions under which ice nucleating particles are formed, and the size and shapes of atmospheric ice crystal are particularly sensitive to the size and chemical composition of seeding aerosol nuclei \citep{burrows2022ice,kanji2017overview,hoose2012heterogeneous}. There are also many open questions regarding the sensitivity of ice formation to thermodynamic conditions, and the interplay between homogeneous freezing and various modes of heterogeneous freezing on a seeding aerosol \citep{burrows2022ice,kanji2017overview}.


Despite decades of laboratory, observational field studies and remote sensing, together with modeling work, fundamental questions remain on the  nucleation and growth of droplets and ice particles on aerosols. These processes play a critical role in atmospheric science, through their influence on cloud processes, weather patterns, and radiative balance of Earth's atmosphere \citep{calvin2023ipcc} and are central to proposed climate mitigation strategies such as marine cloud brightening \citep{feingold2024physical} and cirrus cloud thinning \cite{lohmann2017cirrus}.



Highly controlled laboratory conditions provide a path forward to gain physical insights into the mechanisms responsible for ice and droplet nucleation on seeding aerosols and subsequent growth under various thermodynamical conditions. While experimental cloud chambers have been instrumental in developing understanding of cloud microphysics and informing parameterizations implemented in large scale models\cite{mason1962cloud,schmitt1982university,miller1983homogeneous,pruppacher1998microphysics,shaw2003particle}; only a few chambers are still in operation \cite{shaw2020cloud}. 
As discussed by \citet{shaw2020cloud}, there is a significant need for a renewed effort of experimental studies into cloud and aerosol microphysics in controlled environments (i.e. well-constrained boundary conditions, thermodynamical properties, seeded aerosols, and turbulence conditions). The ideal cloud chamber experiment should include diagnostics capable of detecting multiscale dynamics directly within the flow, including droplet and ice particles size distribution as well as water vapor content and spatial distribution. Also essential are the ability to isolate specific processes and a high degree of repeatability in order to make detailed comparisons between experiments and modeling frameworks. 

Cloud chambers can operate through different working principles. To give two relevant examples, the Aerosol Interaction and Dynamics in the Atmosphere (AIDA) expansion chamber at the Karlsruhe Institute of Technology (KIT)\citep{wagner2009review} and the turbulent $\Pi$ Cloud Chamber at Michigan Technological University\citep{prabhakaran2020role,chang2016laboratory} have contributed to recent scientific discoveries on aerosol cloud microphysics.

The AIDA facility is a large volume (84~m$^3$) expansion chamber capable of studying droplet and ice growth over a range of thermodynamic conditions\citep{wagner2009review}. 
The large chamber size brings certain operational challenges and motivated KIT to develop a smaller vessel, AIDAm, housed inside AIDA. AIDAm (volume of 0.02~m$^3$) is used to study the long-term physical and chemical aging of atmospheric aerosols at relevant atmospheric thermodynamic conditions, and measure the impact of aging on the ice-nucleating properties of the aerosols \citep{vogel2022development}. The Portable Ice Nucleation Experiment (PINE) chamber\citep{mohler2020portable} is another expansion chamber led by KIT, designed as a plug and play facility. PINE is capable of minute-to-hour observations of ice-nucleating particles by controlling the wall temperatures and thereby creating the necessary thermodynamic conditions for ice growth. 
Its small size (volume of 0.01~m$^3$) and portable design permits its deployment in field sites \citep{vogel2024ice} allowing for monitoring ice nucleation on aerosols in various atmospheric backgrounds.

The $\Pi$ chamber operates under Rayleigh-B\'{e}nard turbulent convection conditions to study aerosol-cloud interactions, cloud formation, and turbulence coupling \citep{prabhakaran2020role,chang2016laboratory}. Elegantly, specific humidity and temperature boundary conditions may be chosen to generate supersaturation conditions inside the chamber which lead to droplet nucleation, in contrast to expansion chambers. The $\Pi$ chamber is meter-scaled (volume of 3.14~m$^3$) and operates in a statistically stationary state with constant injection of aerosols to compensate for sedimentation and wall loss. Studies of droplet formation and dynamics over minutes to days have been performed, with a focus on droplet-turbulence interactions with analysis of preferential sampling \citep{shaw1998preferential}, heterogeneous spatial distribution of cloud droplets \cite{kostinski2001scale}, and the effect of fluctuating supersaturation fields in turbulent clouds \cite{prabhakaran2020role}.


Following a community workshop, \citet{shaw2020cloud} laid out a list of lingering open questions in cloud microphysics and the chamber scales needed to address them. While a very tall convection chamber is required for some questions -- rain formation and processes at the cloud boundary where gradients are sharp -- meter scale chambers are sufficient to attack several key questions on aerosol activation, droplet formation and growth, and ice formation and growth. Very small chambers (centimeter-scale) have been used in single-particle growth studies, and they have shown a diversity of ice growth behaviors that may be impossible to directly observe in a large chamber, such as AIDA \citep{bacon2003, pokrifka2020, pokrifka2023}. An intermediate cloud chamber would be more suitable to measure the diversity of ice growth within a population of particles. Smaller chambers also have the advantages of easier logistics and faster experimental repetition rates, making it possible to sweep much wider ranges of conditions. The facility we introduce in this paper has been specifically designed to target such questions.


Here, we present REACh, a new experimental facility based on the principle of an expansion chamber. Our chamber of intermediate volume (0.14m$^3$) includes the possibility to introduce turbulent flow during the expansion (inspired from particles in turbulence studies \cite{bodenschatz2010can,bocanegra2016dispersion,good2014settling}), so that nucleation and growth can be studied in a fluctuating environment. Crucially, our setup allows for a very fast experimental repetition rate, with a typical experiment taking only $\sim$30 minutes to set up and run. We can scan a wide range of aerosol, flow and thermodynamical conditions in just a few days. It is useful to carry out such extensive parameter sweeps with experiments that can go beyond classic atmospheric conditions in terms of particles concentration, level of supersaturation, time scale of expansion, intensity of turbulence, helping to develop more accurate theoretical models. Doing so avoids the extrapolation of parameters when at the edge of typical, or extreme, atmospheric conditions. Fast experimental turnaround times can be intractable in larger chambers. 

Moreover, our chamber design emphasizes a suite of high-frequency, accurate diagnostics to monitor droplet growth and thermodynamical conditions throughout each expansion. Such measurements are essential to advance understanding of cloud microphysics. Nucleation diagnostics are often made post-hoc, i.e. by extracting samples from the chamber and running subsequent analyses. Such methods result in a lack of timescale information needed to constrain particle and cloud growth models. More ideal diagnostics would track the size distribution and number densities of dry aerosols and droplets together with temperature and liquid vapor content \cite{fahey2014aquavit}, and the spatial and temporal fluctuations of these fields. 

Here, we illustrate the capabilities of our rapid-expansion-cloud-chamber on the nucleation of liquid droplets. We make use of time-synchronized, highly temporally resolved measurements to probe the dynamics of the droplet growth for a variety of initial conditions, including saturation ratio, seeding aerosol concentration, and air mixing.

REACh consists of two vacuum chambers: an aerosol-nucleation chamber, where seeding aerosols are introduced and measurements are conducted; and the expansion chamber, into which the aerosol-nucleation chamber can be vented to create the necessary conditions for nucleation.
Detailed quantification of droplet properties is enabled by the use of phase Doppler anemometry (PDA), high-speed in-line holography \cite{fugal2009cloud,erinin2023comparison,beals2015holographic}, and temporally resolved measurements of the chamber temperature and pressure. 
An infrared absorption spectroscopy system to measure saturation ratio is under development and will be presented in a follow-up paper.

The remainder of the paper is organized as follows. We describe the operating principle of the facility, experimental procedures, measurement instrumentation, and facility characterization in \S~\ref{sec:methods}. We illustrate the capabilities of the system by characterizing the thermodynamics and associated droplet nucleation and growth in various aerosol concentration regimes, with dense regimes of droplets characterized by the PDA and dilute regime characterized by the holography in \S~\ref{sec:Results and discussion}. In particular, we characterize how droplet lifetimes, mean diameters, and concentrations vary as a function of initial saturation ratio and aerosol concentration. We provide conclusions and future outlook in Section \S~\ref{sec:conc}.

\section{Experimental Methods}\label{sec:methods}
\begin{figure*}
    \includegraphics{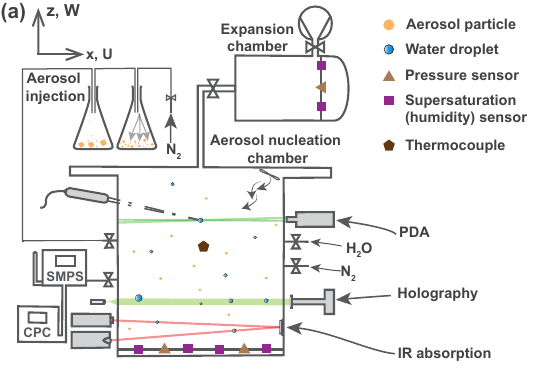}
    \includegraphics{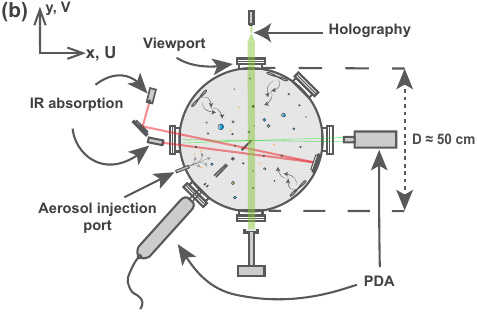}
    \caption{Schematics of the rapid expansion aerosol chamber (REACh). (a) Side view of the aerosol-nucleation and expansion chambers along with instrumentation and control devices, including the aerosol injection system, a scanning mobility particle sizer (SMPS), a condensation particle counter (CPC), a phase Doppler anemometer (PDA) system, an inline holography system, an infrared absorption beamline, pressure sensors, thermocouple temperature sensors, and humidity sensors. (b) Top view of the aerosol-nucleation chamber and the layout of several optical instrumentation systems and the viewports through which they pass into the chamber. The inline holography system can be used to measure droplet concentrations, diameters, and speeds. The partial pressure of H$_2$O vapor can be measured using infrared (IR) absorption spectroscopy.} 
    \label{fig:exp_setup}
\end{figure*}

\subsection{Experimental apparatus and procedure}\label{sec:methods:apparatus}

Figure~\ref{fig:exp_setup} shows a schematic of the rapid expansion aerosol chamber.  
Both the aerosol-nucleation and expansion chambers are depicted as well as various control devices and diagnostics.
The facility consists of two cylindrical vacuum-sealed stainless steel chambers separated by a solenoid valve which is typically in the closed position. 
The primary chamber, hereafter referred to as the aerosol-nucleation chamber, is where all of the experimental measurements are conducted and where most of the scientific instrumentation is located.
The second chamber, hereafter referred to as the expansion chamber, is approximately equal in volume to the aerosol-nucleation chamber. 
We vent the contents of the aerosol-nucleation chamber into the expansion chamber to increase the saturation ratio via adiabatic expansion and cooling, thus  creating conditions for droplet nucleation.

The aerosol-nucleation chamber is 50.8~cm in diameter and 49.5~cm tall with a volume of $\mathcal{V} \approx 0.14$~m$^3$. The volume is slightly greater than that of a cylinder due to the presence of the viewports.
A vacuum sealed access lid is located on top of the chamber, allowing easy access to the main compartment.
The chamber features six 13.5~cm (5.32'') diameter borosilicate glass viewports positioned along its circumference to provide optical access.
In addition, measurement and control devices are attached to several 20.3~cm (8'') and 6.99~cm (2.75'') access ports positioned around the circumference of the aerosol-nucleation chamber and on the top lid.
The chamber walls are equipped with cooling channels which are not used in the present work, but will permit us to explore lower-temperature conditions in the future.
Three 82~mm diameter fans (CUI Devices, model CFM-9225V-130-340) are mounted to the walls of the aerosol-nucleation chamber to enhance mixing inside the chamber. 
The fans are offset by 120$^\circ$ around the circumference of the chamber, positioned near the top and pointing towards the chamber center. 

We produce a cloud of liquid droplets in the aerosol-nucleation chamber using the following procedure.
First, the aerosol-nucleation chamber is prepared with the desired initial saturation ratio, $S_0$, and concentration of seeding aerosols, $C_0$ (see \S~\ref{sec:methods:initial}).
At the same time, the expansion chamber is evacuated to $\sim$2.6$\times$10$^{-5}$~bar using a Roots vacuum pump (Leybold ECODRY 40 plus).
Fully pumping down the expansion chamber results in a maximum pressure drop in the nucleation chamber of $\Delta p \approx 0.54$~bar upon venting; this is the condition used for the majority of experimental runs reported in this paper.
As described in \S~\ref{sec:results:thermo}, we can also work with a smaller pressure drop by only partially evacuating the expansion chamber.

An experiment is initiated when the solenoid valve connecting the two chambers is opened rapidly ($\sim$20~ms opening time), causing gas to flow out of the aerosol-nucleation chamber in $\sim$0.5 seconds.
This expansion drops the aerosol-nucleation chamber pressure, which is accompanied by a drop in temperature and corresponding increase in saturation ratio (defined below).
At this point, water droplets may begin to form in the aerosol-nucleation chamber. 
Nucleation can in principle occur either via heterogeneous nucleation on seeding aerosols or via homogeneous nucleation, depending on the initial conditions prepared in the aerosol-nucleation chamber.
Expansion experiments can be performed without the fans (hereafter referred to as the unforced mixing case) or with the fans running (referred to as the forced mixing case) to evaluate the influence of mixing on droplet formation.

One of the key metrics that governs droplet nucleation in humid air is the saturation ratio, defined as $S = e/e^*(T)$, where $e$ is the partial pressure of water vapor and $e^*(T)$ is the saturation vapor pressure at temperature $T$. $S$ is related to the relative humidity (RH) by $S = \mathrm{RH}/100$. 
A large range of maximum supersaturation ($S-1$) can then be achieved depending on the initial relative humidity, temperature and pressure drop.


For each experimental run the following data are collected: initial temperature, pressure, and saturation ratio; initial aerosol concentration; time resolved pressure and temperature drop; and time-resolved cloud droplet diameter, number, vertical, and horizontal speeds. The chamber pressure, temperature, and saturation ratio are monitored using the following sensors:
\begin{itemize}
    \item Three pressure sensors (Keller Preciseline), two in the aerosol-nucleation chamber and one in the expansion chamber, are used to measure the sudden changes in pressure that occur during an expansion. These sensors feature a sampling rate of 250~Hz and equivalent response times.
    \item One thermocouple temperature sensor (Type T, 44 AWG, Omega Engineering Inc.) is hung in the center of the aerosol-nucleation chamber to measure the temperature drop during an expansion. The estimated response time of the thermocouple is on the order of 10~ms. A cold junction compensation (National Instruments SCB-68A) measuring room temperature, which is held at 20$^\circ$C, is applied to the thermocouple. Note that the thermocouple wire is very fine, thus it moves in the presence of turbulence.
    \item Six relative humidity sensors (Sensirion SEK-SHT35), four near the bottom of the aerosol-nucleation chamber and two in the expansion chamber, are used to record relative humidity. The response time of these sensors is on the order of 2~s, so they are only used to record humidity before and after the expansion.
\end{itemize}
The instrumentation used to record aerosol and droplet statistics is described below in \S~\ref{sec:methods:measurements:aerosols} and \S~\ref{sec:methods:measurements:drops}.

\subsection{Chamber initialization \label{sec:methods:initial}}

We set up for an expansion experiment by initializing the aerosol-nucleation chamber. The most important parameters to control are the initial saturation ratio, $S_0$ and the concentration of dry seeding aerosol particles, $C_0$. It can take 20 to 35 minutes to prepare the chamber, depending on conditions.

The aerosol-nucleation chamber is thoroughly cleaned of aerosols from any previous experiments.
The chamber is then sealed and the room air inside, which contains various uncontrolled aerosols, is displaced with pure nitrogen (N$_2$).
Next, we pass pure N$_2$ through a bubbler filled with ultrapure water from a Milli-Q, creating a mixture of water vapor and droplets suspended in the N$_2$ carrier gas.
The resulting mixture passes through a two-stage compressed gas filter to remove any droplets or particles with diameters above $d > 5$~\textmu{m}, resulting in a stream of humid air. 
This humid air is injected into the aerosol-nucleation chamber at a flow rate of 10 standard liters per minute (SLPM) using a mass flow controller (Alicat Scientific) until the desired $S_0$ is reached.

Once the specified $S_0$ is reached, we set the concentration of dry seeding aerosol particles, $C_0$, in the aerosol-nucleation chamber.
Aerosols are injected into the chamber using a two-stage powder dispersion system, similar to those described by \citet{sullivan2009effect} and \citet{huynh_bottle_system}.
A schematic of the aerosol injection system is shown in the top left of Figure~\ref{fig:exp_setup}(a).
The system consists of two 2~L Erlenmeyer flasks connected in series with the first flask connected to an N$_2$ source and the second to the aerosol-nucleation chamber.
Aerosol powder of known composition is placed in the first flask along with a Teflon-coated stir bar.
The flask is placed on a stir plate set to continuously stir the powder.
A mass flow controller (Alicat Scientific) is used to inject dry N$_2$ into the first flask at a flow rate of 0.4 to 5 SLPM. 
The N$_2$ entrains aerosol particles from the first flask in the flow, then passes into the second flask and eventually into the aerosol-nucleation chamber. 
Sharp turns are introduced into the tubing and inside the second flask to trap large particles thereby enhancing the selection of small particles.

For all experiments in this manuscript, the aerosol-nucleation chamber is seeded  with calcium carbonate particles (CaCO$_3$, $99\%$ pure, American Elements Inc., nominal size of 100~nm). 
We choose calcium carbonate here to demonstrate the capabilities and simple usage of the chamber and associated instrumentation. 
In future experiments, other seeding aerosols relevant to cloud microphysics (e.g.\ mineral dust, sea salt, soot) can be explored to assess the role of hygroscopicity in cloud droplet nucleation \cite{tang2016interactions}.

During initial preparation of the aerosol-nucleation chamber, the fans mounted inside the chamber are kept on to assist with the mixing of humid air and seeding particles.

\subsection{Aerosol measurements}\label{sec:methods:measurements:aerosols}

Two separate measurement systems are used to characterize the seeding aerosols. We use a scanning mobility particle sizer (SMPS) to record size distributions and a condensation particle counter (CPC) to record concentrations.

\begin{figure}
    \begin{center}
        \includegraphics[trim={0.5cm 5.5mm 0 0},clip]{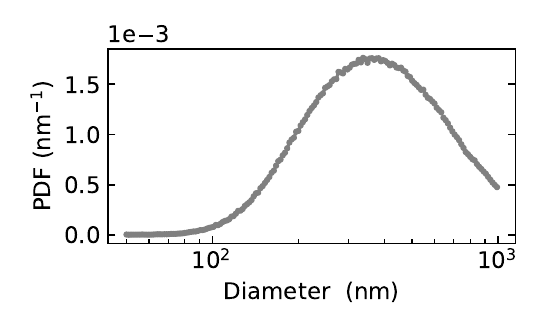}
    \end{center}  
    \caption{Probability density function (PDF) of the diameter distribution of solid CaCO$_3$ aerosol particles introduced into the aerosol-nucleation chamber measured using a scanning mobility particle sizer.}
    \label{fig:size_distribution}
\end{figure}

The SMPS (TSI) is used to characterize the diameter distribution of the dry \ce{CaCO3} particles injected into the chamber, measuring particles sizes between 10~nm and 1~\textmu{m}. The SMPS configuration consists of the elements: advanced aerosol neutralizer (TSI Model 3088); electrostatic classifier (TSI Model 3082); differential mobility analyzer (TSI Model 3081); condensation particle counter (TSI Model 3752).
Figure~\ref{fig:size_distribution} shows the probability distribution function of the CaCO$_3$ particle diameter as measured with the SMPS.
This distribution is centered around $\overline{d} = $ 316~nm with a standard deviation of 257~nm. The difference between the nominal 100~nm particle diameters and the measured diameters is likely due to coagulation of the aerosols during their preparation and seeding.
The shape of the particle size distribution is insensitive to the N$_2$ flow rate used in the powder dispersion system. 

We also use a CPC (TSI model 3007) to measure the concentrations of aerosols up to $10^5$~\#/cm$^3$ with an accuracy of $\pm$ 20 \%.
The CPC is sensitive to particles larger than 10~nm in diameter. 
We use the CPC to sample air from the aerosol-nucleation chamber during the preparation of the experiment to confirm that room aerosols inside the aerosol-nucleation chamber are removed down to a base concentration of $1-2$~\#/cm$^3$. We also use the CPC to record the \ce{CaCO_3} concentration after chamber initialization, moments before an expansion is set to occur.



\subsection{Droplet measurements}\label{sec:methods:measurements:drops}
We characterize the droplets that appear during an expansion using a phase Doppler anemometry (PDA) system (Dantec Inc. Flow Explorer, HiDense FiberPDA optical receiver, and Burst Spectrum Analyzer).
In its present configuration, the PDA can record droplet diameter distributions between 1 and 50~\textmu{m} as well as the horizontal and vertical droplet velocity distributions.
The Flow Explorer consists of two lasers with wavelengths of 532~nm and 561~nm. 
Each beam is split into two; the resulting four beams intersect at a measurement volume of approximately 0.013~mm$^3$ and a projected area of $A$ = 0.119 mm $\times$ 0.1196 mm.
The measurement volume is located close to the center of the aerosol-nucleation chamber, approximately 267~mm from the bottom of the chamber and 155~mm away from the walls.
Horizontal and vertical droplet speeds as well as diameters are measured for all droplets that pass through the measurement volume with a sampling rate of up to 1 MHz.
Detected droplets with diameters smaller than 1~\textmu{m} are removed from the dataset as they are below the measurement accuracy threshold.

In addition, we use a cinematic inline holographic system to measure droplet statistics including size, number, concentration, and speed. A detailed description of this system is provided in \citet{erinin2023comparison}.
We use this holographic technique in \S~\ref{subsec:drop_diam} to report on the temporal evolution of droplet diameters and concentration during an expansion.
In these experiments, the system is recording at a frame rate of 600~Hz and has a spatial resolution of 1.41~\textmu{m} with an ability to measure droplets with diameters down to about 6~\textmu{m}.

The PDA system is designed to operate in relatively dense spray and provides statistics at a single point in space. Meanwhile, the holographic technique probe droplet concentration, sizes, and speed  over an extended pathlength and operates at low droplet concentrations. Holography is especially suited to measurements of ice particles \citep{pasquier2023understanding}, which often nucleate in low concentrations and form nonspherical shapes. In the present paper, we use the PDA for experiments with the larger pressure drop and seeding aerosol concentrations in the range of $10^2$ to $10^5$ particles \#/cm$^3$, which yield droplet concentration of similar orders of magnitude. We use the holographic system for experiments with lower aerosol concentration (typically 1 to 100 particles \#/cm$^3$ and carried out closer to the onset of droplet nucleation. 
We note that the PDA has a much smaller measurement volume than the holographic system, so that the holographic system actually records a much larger number of individual number of droplets than the PDA while operating in much less dense clouds of droplets.


In addition, we record the partial pressure of H$_2$O inside the aerosol-nucleation chamber using infrared laser absorption spectroscopy\cite{buchholz2014rapid,fahey2014aquavit}.
While we do not report the results of these measurements here, we mention this technique here for completeness.



\begin{table*}
    \begin{ruledtabular}
    \begin{tabular}{c c c c c c c c}
        Dataset & Mixing & $S_0$ & $\Delta$ P (bar) & $C_{0}\times$10$^{4}$ (\#/cm$^3$) & \# of runs & Figure \# \\[3pt]
        \hline
        Thermodynamic & Unforced & 0 & 0 -- 0.54 & $\leq 0.0002$ & 11  & 4 \\
        Thermodynamic & Unforced & 0.5 & 0 -- 0.54 & $0.2$ & 11  & 4 \\
        Thermodynamic & Unforced & 0.3 & 0.54 & 0.175 -- 7.05 & 8  & 5 \\
        \hline
       Droplet statistics (PDA) & Unforced & 0.3 & 0.54 & 0.174 -- 24.77 & 7  & 11, 12 \\ 
        Droplet statistics (PDA) & Unforced & 0.5 & 0.54 & 0.141 -- 6.37 & 8  & 11, 12 \\ 
       Droplet statistics (PDA) & Unforced & 0.7 & 0.54 & 0.174 -- 6.49 & 8  & 6, 8 -- 11 \\ 
        Droplet statistics (PDA) & Forced & 0.3 & 0.54 & 0.190 -- 5.03 & 8  & 11 \\ 
        Droplet statistics (PDA) & Forced & 0.5 & 0.54 & 0.165 -- 5.74 & 7  & 11 \\ 
       Droplet statistics (PDA) & Forced & 0.7 & 0.54 & 0.169 -- 6.78 & 7  & 6, 8 -- 10 \\ 
        Droplet statistics (PDA) & Forced/Unforced & 0.7 & 0.54 & 18.5 & 1  & 3, 7 \\ 
        \hline
        Droplet growth (Holography) & Unforced & 0.3 & 0.36 & 0.007 & 1  & 12 \\
    \end{tabular}
    \end{ruledtabular}
    \caption{Summary of the experimental conditions. Thermodynamic measurements are used the characterize the chamber (Section III. A and B). Droplet statistics measurements are collected with PDA for high aerosol concentration (III. C) while droplet growth near nucleation onset are collected with the holographic system for low aerosol concentration (III D). The values of initial saturation ratio $S_0$, \ce{CaCO3} aerosols concentration $C_{0}$ are provided together with the mixing conditions (either unforced fans off or forced fans on).  The figure numbers where each data set is used is provided. Note that $C_0\leq 0.0002$ corresponds to a clean chamber scoured of particles within the detection limit of the CPC. }
    \label{tab:runs_param}
\end{table*}



\section{Results and Discussion} \label{sec:Results and discussion}

We now present the characterization of our cloud chamber in terms of thermodynamical conditions and flow and then demonstrate its capabilities to analyze droplet nucleation for a wide range of initial aerosol concentration and saturation conditions. Table~\ref{tab:runs_param} details all the conditions for experimental results presented in this manuscript.

In \S~\ref{sec:results:thermo} and \S~\ref{sec:results:flow}, we characterize the thermodynamics and flow conditions during expansion experiments for a range of pressure drops and initial saturation ratios, with both forced and unforced mixing. We compare the minimum temperature reached during each expansion to dry and moist adiabatic theories.

In \S~\ref{sec:results:droplets} and \ref{subsec:drop_diam}, we present an analysis of conditions where droplets are formed in presence of \ce{CaCO3} aerosols, for various initial relative humidity (saturation ratios), aerosol concentration and mixing conditions. We leverage PDA (in C) and holography (in D) to probe respectively high and low ranges of aerosol and droplet concentration. 

In \S~\ref{sec:results:droplets}, we present a wide range of $S_0$ condition, with a wide sweep in concentrations of \ce{CaCO3} being performed, for relatively large values of \ce{CaCO3} (from 100 to $10^5$ particles \#/cm$^3$. The droplet statistics are obtained using the PDA. The same data sets are collected for both unforced and forced mixing conditions. For all experiments in \S~\ref{sec:results:droplets}, the initial temperature and pressure are 20.0~$^\circ$C and 1~bar and the pressure drop is always $\Delta p~\approx$ 0.54 bar.


In \S~\ref{subsec:drop_diam}, we report an experiment to demonstrate the use of the holographic system to time resolve droplet growth. Given the large volume sample of holography, this experiment is performed for a small number of aerosol concentration, and we consider a lower pressure drop ($\Delta p~\approx$ 0.36 bar) relatively close to onset of droplet nucleation.





\subsection{Thermodynamics during the expansion} \label{sec:results:thermo}

\begin{figure}
    \centering \includegraphics[width=\linewidth]{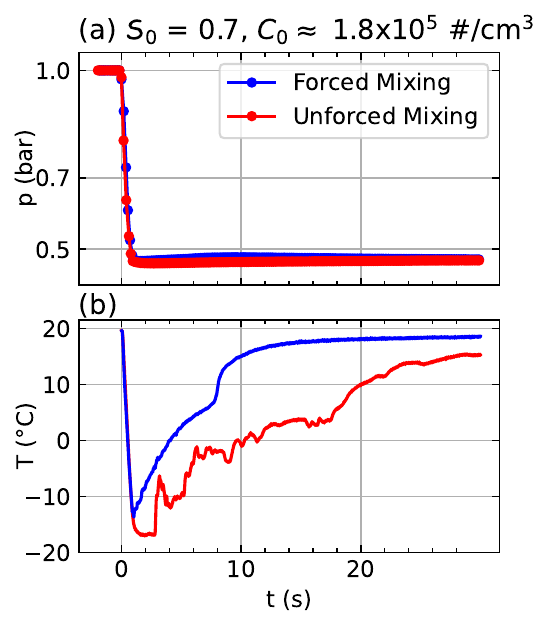} 
    \caption{Time series of (a) pressure and (b) temperature measured during an expansion with $\Delta p = 0.54$~bar, $S_0$ = 0.7, and $C_0 = $ 1.8$\times$10$^{5}$~\#/cm$^3$. 
    Both unforced mixing (red) and forced mixing (blue) cases are shown. The pressure drop is complete in $< 1$~s in both mixing cases. No measurable differences are observed in the initial temperature decrease ($t<0.5$~s), however there is a 3$^\circ$C difference in the minimum temperature, $T_\mathrm{min}$, reached between the two cases. Shortly after $T_\mathrm{min}$ is reached, the chamber begins to thermally equilibrate with the room. This thermalization process is faster in the forced mixing case, when fans induce turbulent mixing.}
    \label{fig:temp_fans}
\end{figure}

Here, we describe measurements of thermodynamic properties inside the chamber throughout an expansion and compare these results to theoretical predictions.
Figure~\ref{fig:temp_fans} shows the typical changes in pressure and temperature during an expansion for an initial saturation ratio of $S_0 = $ 0.7 and an initial \ce{CaCO3} particle concentration of $C_0 = 1.8\times10^{5}$~\#/cm$^3$. Such large concentrations, while not typical of the atmosphere, allow condensation to largely keep up with supersaturation generation during experiments with large pressure drops, keeping the supersaturation close to unity.  
As described in \S~\ref{sec:methods:apparatus}, fans are installed inside the chamber to enhance the mixing of the humid air and aerosols and can be left on during an expansion.
We show results for both fans off (unforced mixing, red curves) and fans on (forced mixing, blue curves) experimental runs.

In both the unforced and forced mixing cases, the rapid pressure decrease lasts approximately 1~s (Figure~\ref{fig:temp_fans}~(a)) and is accompanied by a decrease in temperature ((Figure~\ref{fig:temp_fans}~(b)).
In all experiments, the initial temperature is that of the room, $20~^\circ$C.
Note that in some of the data, the cold junction compensation of $20~^\circ$C was applied in post-processing. Shortly after the expansion, the temperature reaches a minimum, then subsequently increases as the air returns to thermal equilibrium via heat conduction through the chamber walls.
In the unforced mixing case, the minimum temperature reached in the center of the chamber is 3 degrees colder than in the forced mixing case, and persists for $\sim$~3~s, before the air thermalizes over the following $\sim$~30~s; while the thermalization process lasts only $\sim$~15~s in the forced mixing case.
During this thermalization process in unforced conditions, the temperature fluctuates by $\pm5~^\circ$C, likely due to large scale convective air currents inside the chamber.
Thermalization is faster in the forced mixing case as the fans enhance the thermal conduction of heat through the air-wall boundary.

In both experiments shown in Figure~\ref{fig:temp_fans}, a sudden temperature increase is observed during the thermalization process, occurring $\sim17$~s after the expansion is triggered in the unforced mixing case and after $\sim6$~s in the forced mixing case.
These times correspond approximately to the cloud lifetime, $\tau_f$, which we define as the length of time over which cloud droplets are present and detectable after an expansion.
Measurements of the cloud lifetime are discussed in more detail in \S~\ref{subsubsec:influence_of_fans}. 
Eventually, the cloud of droplets evaporates; thermalization can then accelerate afterwards, as the heat flux from the chamber walls goes entirely towards sensible heating (raising the temperature), rather than providing latent heat for droplet evaporation. 
This process explains the sudden changes observed in temperature as the chamber returns to thermal equilibrium.

 \begin{figure}
    \begin{center}
        \includegraphics[scale=0.85]{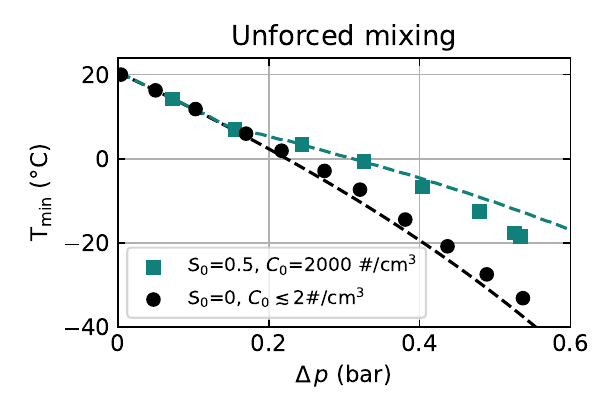}
    \end{center}
    \caption{Minimum temperature reached during expansions as a function of the pressure drop $\Delta p$.  Data is shown with unforced mixing for dry conditions ($S_0 = 0$) with minimal seeding aerosol particles and moist ($S_0 = 0.5$) conditions with $C_0=2000$~\#/cm$^3$. The markers correspond to experimental data while the dashed curves show the relations Eq.\ \eqref{laplace_law} in the dry case and Eq.\ \eqref{eq:moist} in the moist case.}
    \label{fig:temp_min}
\end{figure}

The minimum temperature reached in the aerosol-nucleation chamber during an expansion, $T_\textrm{min}$, is a key thermodynamic variable and is an important determinant of the total condensed water after expansion.
In Figure~\ref{fig:temp_min}, we report $T_\textrm{min}$ for a series of pressure drops with $\Delta p = p_f - p_0$ ranging from $0.02$ to $0.54$~bar.
Figure~\ref{fig:temp_min} shows results for both the dry expansion case with $S_0= 0$ (black circles) and minimal seeding aerosols, and the moist expansion case with $S_0 = 0.5$ (teal squares) and $C_0=2000$~\#/cm$^3$.
In both cases, the minimum temperature decreases as $\Delta p$ increases. 

The dry expansion experiments can be modeled as a dry adiabatic expansion, with \cite{yau1996short} 
\begin{equation}
    T_{f} = T_0 \left( \frac{p_{f}}{p_0} \right)^{\frac{\gamma - 1}{\gamma}}
    \label{laplace_law}
\end{equation}
where $T_0$ and $p_0$ are the initial temperature and pressure, $T_f$ and $p_f$ are the final temperature and pressure after the expansion, and $\gamma = C_p / C_V$ is the ratio of specific heats for dry nitrogen.  $T_f$ can be directly replaced by $T_{\mathrm{min}}$. 
The black dashed line in Figure~\ref{fig:temp_min} shows the expected $T_\mathrm{min}$ using Eq.~\eqref{laplace_law} with the measured values of $p_0$, $T_0$ and $p_f$ and assuming $\gamma = 1.4$. This calculation shows good agreement with the experimentally measured $T_\mathrm{min}$.

The maximum possible saturation ratio $S_{max}$ during an expansion can be estimated from the pressure drop and the estimated minimum temperature from the dry adiabat, $S_{max}=e/e^*(T_{min})$. Assuming adiabatic expansion \citep{miller1983homogeneous}, the water vapor pressure is estimated as $e=e_0\times(p_f/p_0)$ where $e_0$ is the initial water vapor pressure (given by $S_0$), $p_0$ is the initial total chamber pressure, and $p_f$ is the final chamber pressure after expansion. We obtain that for the highest pressure drop $\Delta p \approx 0.54$, $S_{max}$ is up to 10, while for $\Delta p \approx 0.42$, $S_{max}\approx 4$ and for $\Delta p \approx 0.23$, $S_{max}\approx 1$. These values are calculated assuming an initial $S_0$ of 0.3.  We note that the highest $S_{max}$ values are not actually happening in the chamber as water vapor condense on seeding aerosols, but rather indicate that the range of pressure drop available creates conditions strongly conducive to nucleation and condensation.


Most of our expansion experiments are carried out in a humid environment, where Eq.~\eqref{laplace_law} is no longer valid. 
In order to validate the thermodynamics of the expansions at these conditions, we must modify to Eq.~\eqref{laplace_law} to account for humidity.
Determining \Tf\ as a function of \pf\ for humid air with $0<\RH<1$ requires two ingredients. First, we must determine the temperature \Tsat\ and pressure \psat\ at which the expanding humid chamber air initially reaches saturation ($S = 1$), and second, we must generalize Eq.~\eqref{laplace_law} for saturated air. 

We begin by noting the following relations where $e$ denotes the partial pressure of water vapor and $e^*(T)$ is the partial pressure of water vapor at saturation for temperature $T$: 
\begin{align}
    \left(\frac{\Tsat}{\To}\right)^{\Cp/\Rd} & = \frac{\psat}{\po} \label{psat}\\
        & = \frac{e_{\text{sat}}}{e_0} \\ 
        & = \frac{e^*(\Tsat)}{\RH \, e^*(\To)} \\
        & = \frac{1}{\RH}\exp\left[-\frac{L_v}{\Rv}\left(\frac{1}{\Tsat}-\frac{1}{\To}\right)\right] \label{C-C} 
\end{align}
The first equality in the expressions above follows from Eq.~\eqref{laplace_law}; the second follows from the fact that the partial pressure of water vapor scales with the total pressure in the absence of condensation; the third arises from the definitions $e_\textrm{sat} = e^*(T_\textrm{sat})$ and $S_0 = e_0/e^*(T_0)$; and the last derives from the Clausius-Clapeyron relationship for $e^*(T)$, namely $de^*/dT=-(L_v/\Rv T^2)e^*(T)$.\citep{rogers1989}
Note that \Rd\ and \Rv\ are the gas constants for $N_2$ and water vapor, respectively and $L_v$ is the latent heat.

Similar to the manipulations performed by \citet{romps2017}, we solve the above equation for \Tsat\ in terms of \RH\ using the `$-1$' branch of the Lambert W-function which satisfies  $W(xe^x)=x$, obtaining
\begin{equation}
    \Tsat \ = \ \frac{-c_i }{W_{-1}\left(-\RH^{\Rd/\Cp}c_i e^{-c_i}\right)}\To
    \label{Tsat}
\end{equation}
where $c_i\equiv \Rd L_v/(\Cp\Rv \To)$.
Eq.~\eqref{Tsat} can then be combined with Eq.~\eqref{psat} to obtain an equation for \psat:
\begin{equation}
    \psat \ = \ \po \bigg[ \frac{-c_i}{W_{-1}\left(-\RH^{\Rd/\Cp}c_i e^{-c_i}\right)} \bigg]^{\Cp/\Rd} \ .
\end{equation}
Note that with \Tsat\ and \psat\ determined, we can then determine the specific humidity \qsat\ (kg \ce{H2O}/kg humid air) at the point of first saturation  in terms of the saturation specific humidity $q^*$ evaluated at $(\Tsat,\psat)$:
\begin{equation}
    \qsat \ = \ q^*(\Tsat,\psat) \ = \ \frac{\Rd}{\Rv}\frac{e^*(\Tsat)}{\psat} \ .
    \label{qstar}
\end{equation}

We note that because S might exceed 1 in the chamber, $T_{sat}$ will be an upper bound of the minimal temperature while T$_{min}$ obtained from the dry adiabat is a lower bound.

With \Tsat, \psat, and \qsat\ in hand,  we now seek a relation between $T$ and $p$ after the parcel saturates, assuming that condensation occurs sufficiently rapidly to keep $S$ close to 1 (i.e. $S-1\sim \mathcal{O}(0.01)$) throughout the process. This requires accounting for the effects of the latent heat of condensation on the parcel as it expands and cools. To proceed analytically, we utilize the formula for $q^*(T)$ along such a parcel trajectory as derived in \citet{romps2016clausius}. This formula, unlike the more general formula Eq.~\eqref{qstar}, gives $q^*$ as a function of $T$ along a moist adiabat with initial temperature \Tsat\ and initial specific humidity \qsat, where the $p$-dependence of $q^*$ appearing in \eqref{qstar} is implicitly accounted for using  moist adiabatic thermodynamics. The result is somewhat involved, but can again be expressed analytically in terms of the Lambert W-function:
\begin{align}
    q^*(T) & \ = \ \frac{\Rd\To}{L_v}W_{-1}\left(\ysat \, e^{-f(\Tsat-T)}\right) \label{qvstar_romps} \\
    \mbox{where} \quad \ysat\ & \ \equiv \ \frac{L_v q^*(\Tsat)}{\Rd\To}\exp\left(\frac{L_v q^*(\Tsat)}{\Rd\To}\right) \notag \\
    \mbox{and} \quad  f & \ \equiv \frac{L_v}{\Rv \To^2} - \frac{\Cp}{\Rd\To} \ .  \notag
\end{align}
Evaluating Eq.~\eqref{qvstar_romps} at the final temperature \Tf\ and combining with Eq.~\eqref{qstar} allows us to solve for the final pressure \pf\ in terms of the final temperature \Tf:
\begin{equation}
    \pf \ = \ \frac{\Rd}{\Rv}\frac{e^*(\Tf)}{q^*(\Tf)} \label{eq:moist}
\end{equation}
This yields the desired relationship between the final (minimum) temperature \Tf\ and the pressure drop $\Delta p = \po-\pf$. 

Figure 4 compares the measured values of $T_\textrm{min}$ to those predicted by Eq.\ \eqref{eq:moist} for a representative humid expansion (teal trace and markers). We observe a good agreement between data and theory for pressure drops up to $\Delta p \approx 0.45$. 

\begin{figure}
    \centering \includegraphics[width=\linewidth]{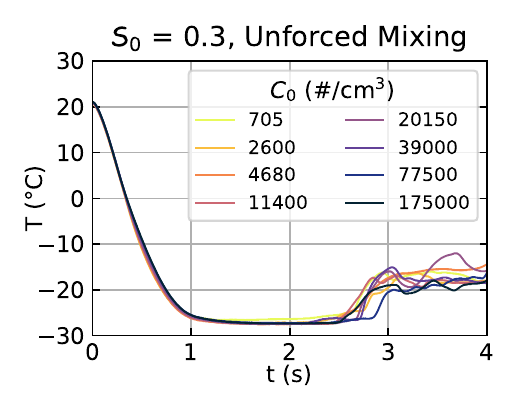} 
    \caption{Time series of temperature during expansions for different initial aerosol concentrations, $C_0$, performed with unforced mixing, an initial saturation ratio of $S_0 = 0.3$, and a pressure drop of $\Delta p \approx $ 0.54~bar. The expansion is triggered at $t = 0$~s. No notable dependence of the temperature curves on $C_0$ are observed during the expansion, from $t = 0$ to 2.5~s. For $t > 2.5$~s, differences in the temperature response are likely due to post-expansion air mixing .}
    \label{fig:temp_min_c0_span}
\end{figure}

Finally, we characterize how the expansion thermodynamics are impacted by the presence of seeding aerosols in various concentrations at the highest pressure drop. 
Figure~\ref{fig:temp_min_c0_span} shows the time evolution of temperature during expansions with different aerosol concentrations, $C_0$. These experiments are carried out with unforced mixing, an initial saturation ratio of $S_0 = 0.3$, and the maximum pressure drop of $\Delta p =$ 0.54~bar. 
The dependence of $T_{\mathrm{min}}$ on $C_0$ is found to be very weak, with variations of at most 0.5~$^\circ$C.
After the expansion, for times $t > 2.5$~s, random temperature fluctuations are observed, which vary from run to run.
We note that this insensitivity of $T_\textrm{min}$ to $C_0$ is likely due to the relatively large aerosol concentrations, which keep the  saturation ratio  not too far from 1 despite the large pressure drop. 

\subsection{Flow in forced and unforced mixing conditions}\label{sec:results:flow}

\begin{figure}
    \begin{center}
        \includegraphics[trim={0.4cm 4mm 0 0},clip]{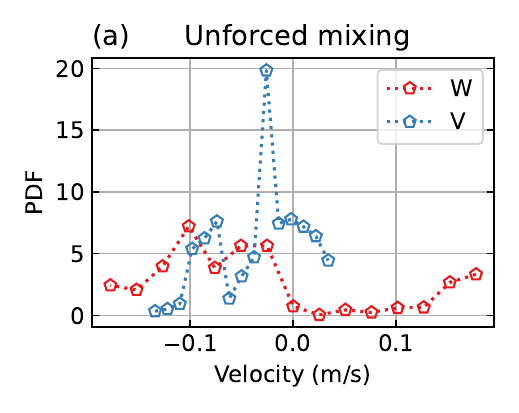} \\
        \includegraphics[trim={0.4cm 4mm 0 0},clip]{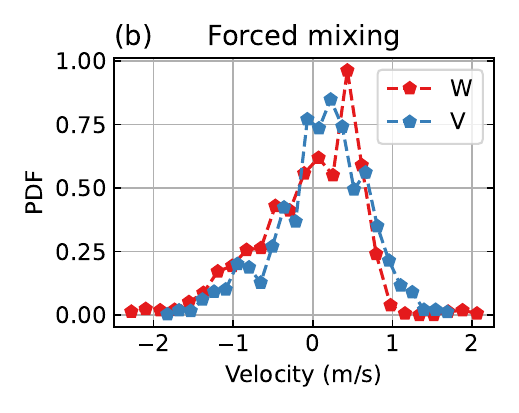} 
    \end{center}
    \caption{Probability distribution functions (PDFs) of droplet velocities recorded with phase Doppler anemometry (PDA) during expansions with (a) unforced mixing conditions (dotted lines) and (b) forced mixing conditions (dashed lines). The velocity $V$ corresponds to the horizontal component of the velocity field, and $W$ corresponds to the upward vertical component in the plane orthogonal to the PDA laser beams; see Figure \ref{fig:exp_setup} for a graphical orientation of the axes. Experiments are conducted with $S_0$ = 0.7 and $C_0 = 8\cdot10^4$~\#/cm$^3$ and $\Delta p = $ 0.54~bar.}
    \label{fig:flow}
\end{figure}

We now characterize mixing inside the chamber by comparing droplet velocity statistics acquired with phase Doppler anemometry for unforced and forced mixing conditions. The distribution and statistics are collected over the full time of the experiment.

Figure~\ref{fig:flow} shows representative horizontal ($V$) and vertical ($W$) velocity probability density functions (PDFs) for both mixing cases with $S_0 = 0.7$ and $C_0 = 8\times10^4$~\#/cm$^3$.
In the unforced mixing case shown in Figure~\ref{fig:flow}~(a), the droplets feature horizontal and vertical velocity distributions that are not centered about zero, and do not have clearly defined shapes due to the limited statistics. 
In the forced mixing case shown in Figure~\ref{fig:flow}~(b), a random turbulent flow is induced by the motion of the fans. Both $V$ and $W$ have peaks at $\sim 0.2$~m/s. The $V$ and $W$ distributions display absolute values nearly an order of magnitude larger in the forced mixing case than in the unforced mixing case. Velocity distributions are very similar across the various $C_0$, $S_0$ and $\Delta p$ conditions tested.

We define the mean droplet speed as  $\bar{U} = \overline{\sqrt{V^2+W^2}}$ where $V$ and $W$ are the measured velocities by the PDA and the overline indicates the average (over drops passing through the sensor over time), and a characteristic fluctuating velocity as $\mathcal{U}={\sqrt{\sigma_V^2+\sigma_W^2}}$ where $\sigma$ represents the standard deviation. 
The mean drop speed in forced conditions is $\bar{U}_{\mathrm{forced}}= 0.8$~m/s and in unforced conditions $\bar{U}_{\mathrm{unforced}}= 0.1$~m/s (with variations of less than 0.1 and 0.01 across conditions respectively). The characteristic velocity fluctuations are very similar so that either can be used as characteristic velocity in the rest of the paper. 



The mixing in the chamber can be dominated either by diffusion or by turbulent mixing; this distinction is characterized by the P\'{e}clet number, $\mathcal{P}_\mathrm{e}$.
$\mathcal{P}_\mathrm{e}$ is defined as the ratio of the diffusion timescale, $\tau^{\mathrm{diff}} = L^2 / \alpha$, divided by the advective turbulent mixing timescale, $\tau^{\mathrm{adv}} = L / \bar{U}$, where $L$ and $\bar{U}$ represent the characteristic length-scale and speed of the flow and $\alpha$ represents the thermal diffusivity of the gas inside the chamber.
We can therefore express the P\'{e}clet number as 
\begin{equation}
    \mathcal{P}_\mathrm{e} = \frac{\tau^{\mathrm{diff}}}{\tau^{\mathrm{adv}}} = \frac{L \bar{U}}{\alpha}
\end{equation}
%
Taking $L~=~0.5$ m, $\alpha = 1.9 \times 10^{-5}$ m$^2$/s taken at $0^\circ$C, and $\bar{U}$, the average droplet speed, the order of magnitude of the P\'{e}clet numbers for the unforced and forced mixing cases are $\mathcal{P}_{\mathrm{e,unforced}} = 2.7\times 10^{3}$ and $\mathcal{P}_{\mathrm{e,forced}} = 2.2\times10^{4}$. 
Both cases feature $\mathcal{P}_\mathrm{e} \gg 1$ and are therefore dominated by mixing.
The difference in the mixing timescales between the two cases can be estimated by:
\begin{equation}\frac{\tau_{\mathrm{unforced}}^{\mathrm{adv}}}{\tau^{\mathrm{adv}} _{\mathrm{forced}}} = \frac{\bar{U}_ {\mathrm{forced}}}{\bar{U}_ {\mathrm{unforced}}}  \approx 8
\end{equation}
Mixing is about 8 times faster in the forced mixing case, so that the fans will enhance the chamber thermalization after an expansion.


\begin{figure*}
    \centering
    \includegraphics[height=8in]{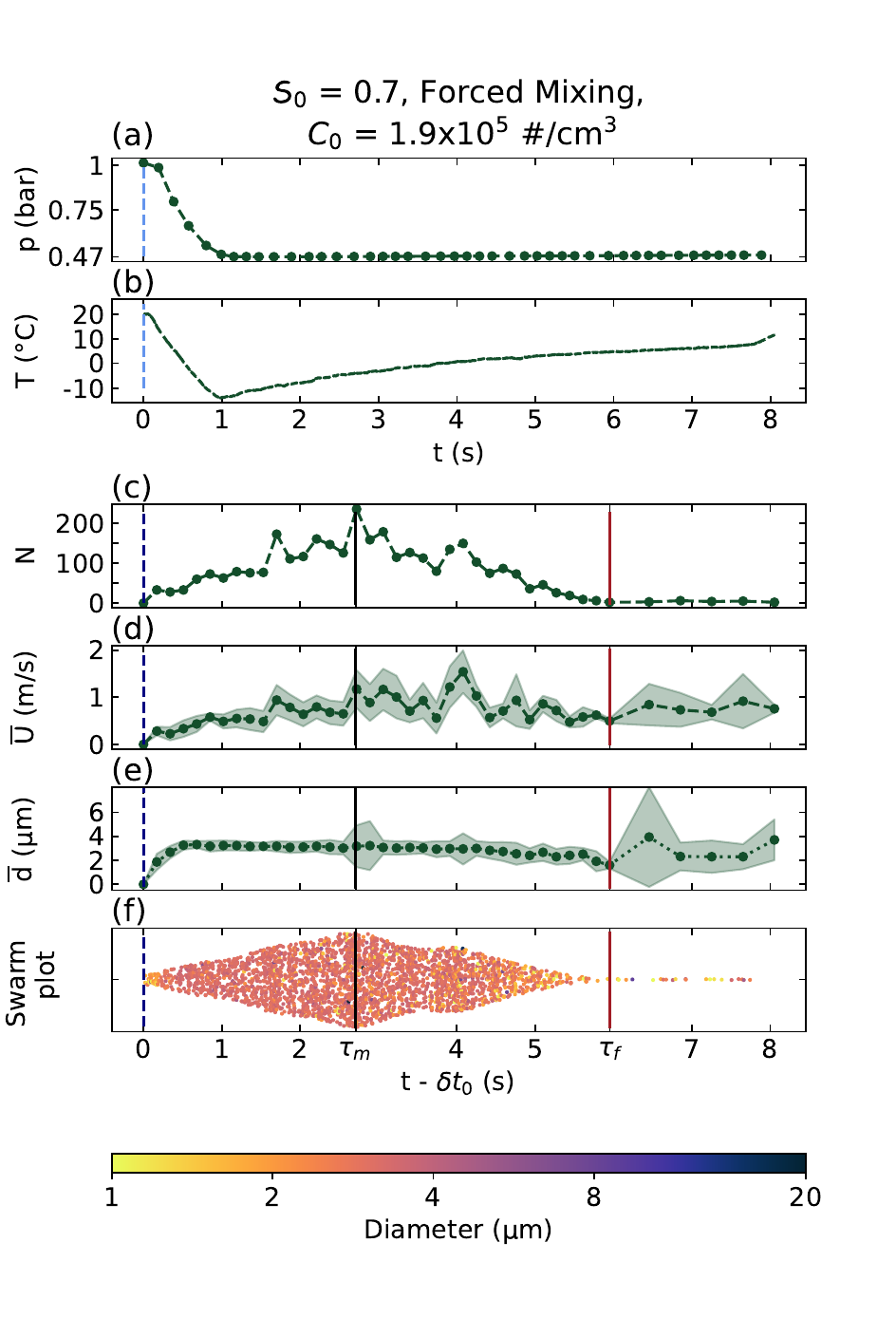}
    \caption{Time series of (a) pressure, (b) temperature, (c) number of droplets detected per 0.17~s time interval, $N$, (d) mean droplet speed $\bar{U}$, (e) running mean droplet diameter, $\overline{d}$, and (f) a swarm plot of individual droplet sizes. Subplots (c-f) show data collected with PDA. All data is collected in real time for a single expansion with $\Delta p = $ 0.54~bar and forced mixing. The initial conditions are $S_0 = 0.7$, $C_0$ = 1.9$\times$10$^{5}$~\#/cm$^3$. The mean droplet speed $\bar{U} = \overline{\sqrt{V^2+W^2}}$ and $\overline{d}$ are calculated over time interval of 0.17~s. The green contours around the $\bar{U}$ and $\overline{d}$ curves represents a $\pm1$ standard deviation. The swarm plot in subplot (f) provides a compact representation of the droplet number and size shown in subplots (c) and (e): each point represents one detected droplet. The evolution of the number of particles is tracked through the thickness of the swarm plot on the vertical axis. The color of each point represents the diameter of that droplet according to the color bar at the bottom of the figure.  
    The light blue dashed line in subplots (a-b) represents the moment the valve opens and the expansion begins. The black dashed line in subplots (c-f) marks the time at which  the first droplet is detected. The exact time difference between these two events, $\delta t_0$, is on $\sim 0.7$s.
    $\tau_m$ and $\tau_f$ denote the middle and end of the cloud lifetime and are indicated by the solid black and red lines in subplots (c-f).}
    \label{fig:time_series}
\end{figure*}

\subsection{Droplet nucleation as a function of seeded aerosols, saturation and mixing \label{sec:results:droplets}}

We now characterize droplet nucleation as a function of initial aerosol concentration and saturation ratio.
We perform a parameter sweep over 46 conditions while tracking the size and number of droplets formed throughout the expansion. All experiments are performed at the maximum pressure drop ($\Delta p =$ 0.54~bar). 

\subsubsection{Droplet nucleation and growth} 

We start by discussing the droplet nucleation dynamics for a typical case.
Figure~\ref{fig:time_series} shows the time evolution of the thermodynamic chamber parameters alongside droplet statistics measured with PDA from the time the pressure drop is initiated ($t = 0$~s) to shortly after the majority of droplets disappear ($t \approx 8$~s). This data is from a single representative expansion with $S_0 = 0.7$ and $C_0$ = 1.8$\times$10$^{5}$~\#/cm$^3$ under forced mixing conditions.
As discussed in \S~\ref{sec:results:thermo}, the sharp drop in pressure (Figure~\ref{fig:time_series}~(a)) triggers a rapid temperature decrease (Figure~\ref{fig:time_series}~(b)).
The drop in temperature causes a sudden increase in saturation ratio which initiates nucleation of water droplets. 

Figure~\ref{fig:time_series}~(c) plots the time evolution of $N$, the number of droplets detected  per 0.17~s time interval, beginning shortly after expansion occurs. 
$N$ increases until time $\tau_m$, as marked with a  black vertical bar in Figure~\ref{fig:time_series}~(c-f).
The cloud lifetime,  $\tau_f$, is the point after which very few droplets are detected as marked with a red vertical bar in Figure~\ref{fig:time_series}~(c-f).

The temporal evolution of the mean droplet speed, $\bar{U}$ (see III.B), is shown Figure~\ref{fig:time_series}~(d). Over the time window $0 < t < \tau_m$, the mean droplet speed is found to increase from $\sim$0.2 to 1.1~m/s. This increase correlates with the time evolution of $N$.
After $t > \tau_m$, the mean droplet speed increase is sustained for several more seconds until $t \sim 4$~s after which it decreases as $t$ approaches $\tau_f$ and the cloud dissipates.

The time evolution of the mean droplet diameter, $\overline{d}$ is shown in Figure~\ref{fig:time_series}~(e). During the initial stages of droplet growth, $\overline{d}$ increases from $\sim1$~\textmu{m} near $t = 0$~s to $\sim4$~\textmu{m} at $t=0.5$~s, then stays nearly constant until $t = \tau_f$.

The swarm plot in Figure~\ref{fig:time_series}~(f) illustrates each detected droplet as a point. This plot captures both the evolution of droplet number, in the vertical width of the swarm, and droplet diameter, as indicated by the color of each point.
After $t > \tau_f$ most droplets have evaporated.

A notable feature of Figure \ref{fig:time_series} is that the droplet diameter \dbar\ is fairly constant between $\tau_m$ and $\tau_f$, while droplets are evaporating and $N$ is decreasing. This is consistent with the ``inhomogenous mixing'' paradigm \citep{latham1977,baker1980}, where the time it takes thermodynamic cloud properties (such as temperature) to homogenize is much longer than the time it takes droplets to equilibrate with their local environment. In this paradigm, heat transferred from the chamber walls completely evaporates the droplets only in nearby pockets of air, rather than rapidly mixing throughout the volume and partially evaporating droplets throughout. This leads to little change in \dbar\ but a change in the average $N$ passing through the PDA detection volume. These results support recent reports suggesting that inhomogenous mixing is a rather ubiquitous phenomenon in cloudy air masses. \citep[e.g.][]{beals2015,hoffmann2019,yeom2023cloud} 


\subsubsection{Cloud lifetime}\label{subsubsec:influence_of_fans}

\begin{figure}
    \centering
    \includegraphics[trim={0 2mm 0 0},clip]{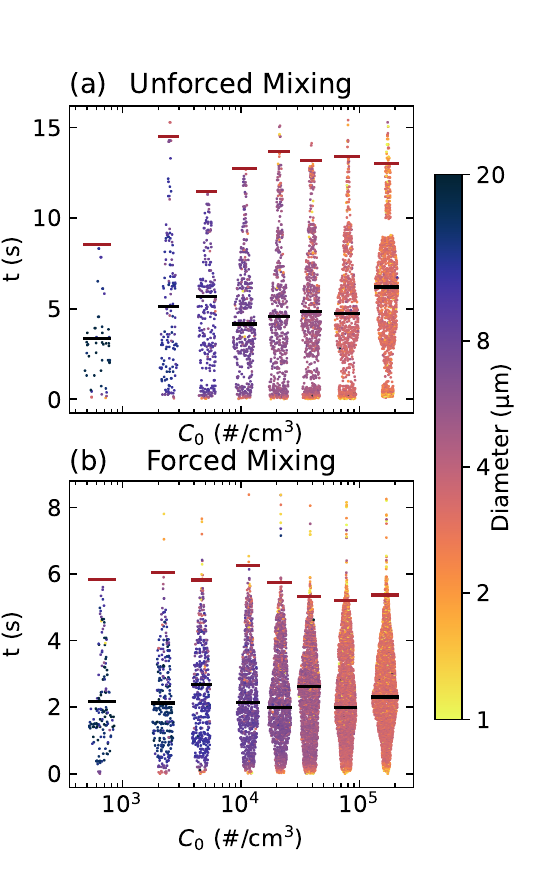}
    \caption{Swarm plots summarizing the time-dependent evolution of droplet number and diameter while sweeping initial aerosol concentration, $C_0$, for (a) unforced and (b) forced mixing with $S_0$ = 0.7 and $\Delta p = $ 0.54~bar. The swarm plots use droplet data from the PDA to show droplet counts, indicated by dots, and diameter, indicated by dot colors. The black and red horizontal bars represent $\tau_m$ and $\tau_f$, respectively. Droplets colored in yellow have measured diameters between $d =  0.1$ and 1~\textmu{m}.  }
    \label{fig:swarm_plot}
\end{figure}

We now characterize the droplet cloud lifetime, $\tau_f$, as a function of the initial saturation ratio, initial aerosol concentration, and mixing conditions.
Figure~\ref{fig:swarm_plot} shows several swarm plots illustrating the time-dependent number and size distribution of droplets for various initial aerosol concentrations, $C_0$. For the smaller values of $C_0$ tested, droplet characterization becomes challenging as very few droplets are measured.
An initial saturation ratio of $S_0$ = 0.7 is used for all experiments reported in Figure~\ref{fig:swarm_plot}.

In the unforced mixing cases shown in Figure~\ref{fig:swarm_plot}~(a), the number of droplets gradually grows to a maximum around $\tau_m = 5-6$~s, as marked with black horizontal bars, and decays to nearly zero by $\tau_f = 10-15$~s, as marked with red horizontal bars.
In the forced mixing cases in Figure~\ref{fig:swarm_plot}~(b), the peak droplet numbers are instead reached at $\tau_m = 2-3$~s, while droplets disappear by $\tau_f =6$~s.
The droplet cloud lives less than half as long under forced mixing than in the unforced mixing case; again, the fans speed the thermalization process. 
As in Figure \ref{fig:time_series}, the drop diameter is fairly constant in time while the droplet number decreases between $\tau_m$ and $\tau_f$, again indicative of inhomogeneous mixing.


%

\begin{figure}
    \centering
        \includegraphics[trim={0 6mm 0 0},clip,width=0.485\textwidth]{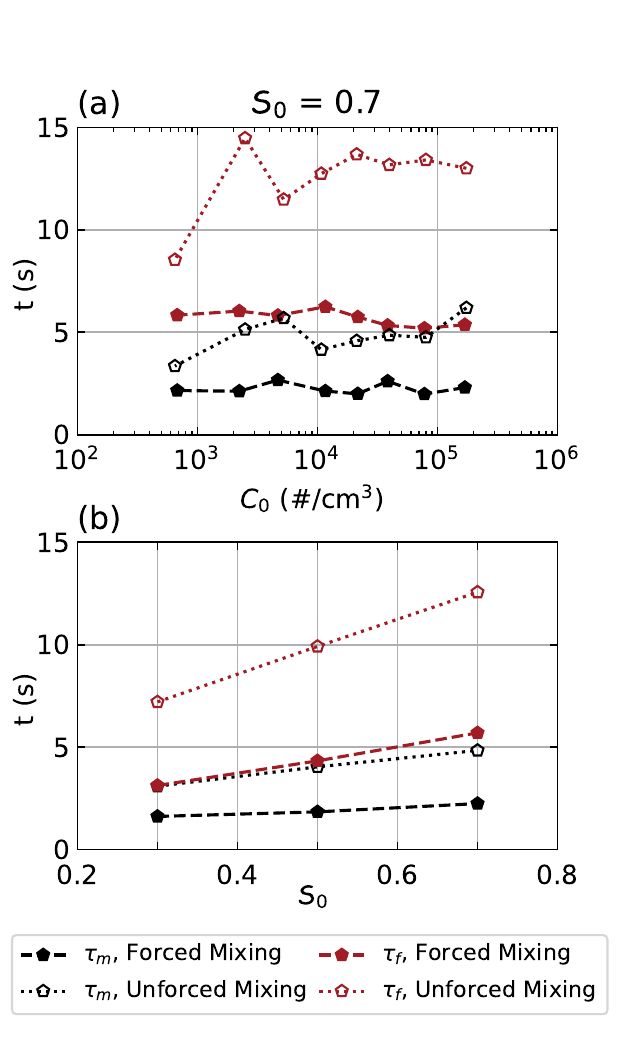}
    \caption{Characteristic timescales of the droplet cloud, $\tau_m$ (black points) and $\tau_f$ (red points), as a function of (a) initial aerosol concentration $C_0$ and (b) saturation ratio, $S_0$. Data is shown for both unforced mixing (empty markers) and forced mixing (solid markers). The marker shapes in subplot (b) are consistent with the markers used for different values of $S_0$ in Figures \ref{fig:stats} and \ref{fig:rescale_concentration}. In subplot (a), $S_0  = 0.7$, while $C_0$ is swept. The timescales plotted in subplot (b)  are calculated by averaging over all concentrations for each value of $S_0$.}
    \label{fig:lifetime_70rh}
\end{figure}

Figure~\ref{fig:lifetime_70rh}~(a) shows the extracted $\tau_m$, and $\tau_f$ timescales as a function of $C_0$ from the same experimental data reported in Figure~\ref{fig:swarm_plot}. $\tau_m$ and $\tau_f$ appear to be independent of $C_0$. The only notable trend is that all timescales are shorter with forced mixing, as already noted above. Meanwhile, Figure \ref{fig:lifetime_70rh}~(b) shows $\tau_m$ and $\tau_f$ as a function of $S_0$. Both $\tau_m$ and $\tau_f$ increase as $S_0$ increases.
As we will see in \S~\ref{subsec:statistics} below, higher initial water content facilitates the growth of droplets to larger average diameters, which can then persist for a longer period of time before evaporating.

\subsubsection{Scaling of droplet statistics}\label{subsec:statistics}

\begin{figure*}
    \begin{tabular}{c c}
        \includegraphics{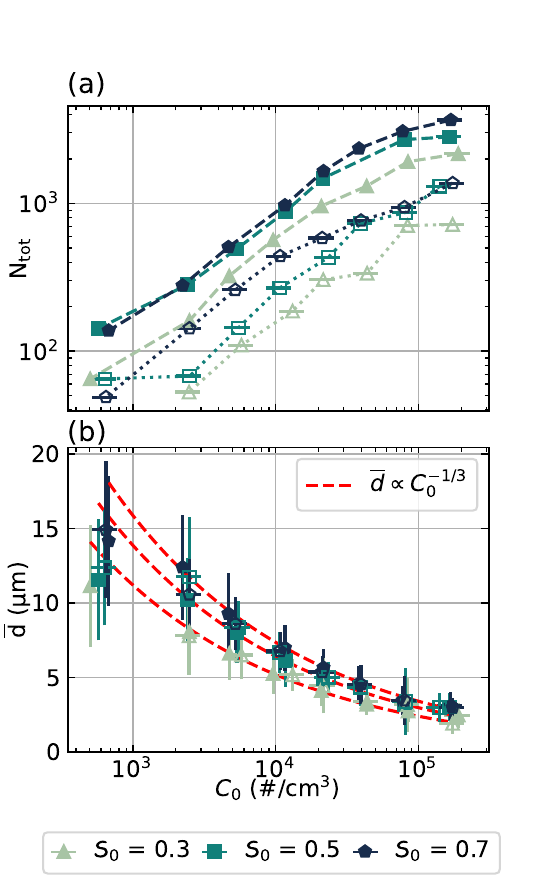} &
        \includegraphics{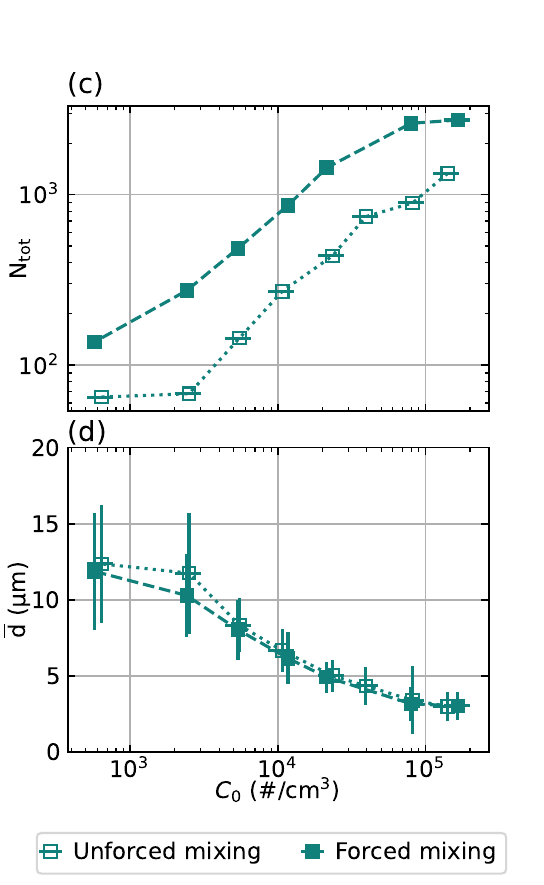}
    \end{tabular}
    \caption{Droplet statistics as a function of initial aerosol concentration, saturation ratio, and mixing conditions. (a) The total number of droplets detected, $N_{\text{tot}}$, and (b) their mean diameter, $\overline{d}$, as a function of $C_0$ for $S_0$ = 0.3, 0.5, and 0.7 and both unforced and forced mixing. (c) $N_{\text{tot}}$ and (d) $\overline{d}$ as functions of $C_0$ for $S_0$ = 0.5. The vertical error bars show $\pm$1 standard deviation in measurement error and the horizontal error bars show measurement uncertainty in $C_0$. Data for $S_0$ = 0.3, 0.5, and 0.7 are plotted with triangle, square, and pentagon markers, respectively, while unforced and forced mixing cases are plotted with empty and filled markers, respectively. The red dashed lines show the $\overline{d}\propto C_0^{-1/3}$ power-law scaling given in Eq.~\eqref{eq:size_vs_c0}.} 
    \label{fig:stats}
\end{figure*}

We now analyze the total number of nucleated droplets and their mean sizes as a function of the initial aerosol concentration $C_0$ and saturation ratio $S_0$. We explored calculating the total number of drops by integrating $N$ over either $0<t<\tau_m$ or $0<t<\tau_f$. and both give similar trends as a function of $C_0$ and $S_0$. The mean droplet diameter is also nearly identical when evaluated over both time windows. We therefore define $N_\textrm{tot}$ as the total number of droplets detected by the PDA over $0<t<\tau_f$ and $\overline{d}$ the average diameter over the same window as counting more droplets yields better statistics. 

Figure~\ref{fig:stats}~(a) plots the dependence of the total droplet number, $N_\mathrm{tot}$
on $C_0$ and $S_0$ for both unforced and forced mixing.
As $C_0$ increases, $N_\mathrm{tot}$ increases; this trend remains consistent across all $S_0$ and mixing conditions. 
$N_\mathrm{tot}$ is also observed to
increase with $S_0$. The relationship between $N_\mathrm{tot}$ and $S_0$ is related to the longer droplet cloud lifetime (see Figure \ref{fig:lifetime_70rh}~(b)).

Figure~\ref{fig:stats}~(b) plots the dependence of the mean droplet diameter, $\overline{d}$, 
on $C_0$ and $S_0$ for both unforced and forced mixing.
$\overline{d}$ consistently decreases with increasing $C_0$ and increases with increasing $S_0$. 

\begin{figure}
    \centering    \includegraphics{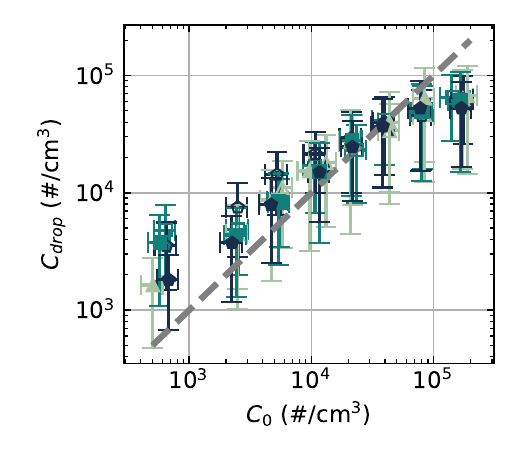}
    \caption{Droplet concentration, $C_{\mathrm{drop}}$, as a function of initial aerosol concentration, $C_0$, for all experimental conditions. Data for $S_0$ = 0.3, 0.5, and 0.7 are plotted with triangle, square, and pentagon markers, respectively, while unforced and forced mixing cases are plotted with empty and filled markers, respectively.  $C_{\mathrm{drop}}$ is calculated according to Eq.~\ref{eq:cdrop}. The gray dashed line has a slope of $C_\textrm{drop}/C_0=1$.}
    \label{fig:rescale_concentration}
\end{figure}

Figure \ref{fig:stats}~(c) highlights how $N_\mathrm{tot}$ and $\overline{d}$ compare between unforced and forced mixing for $S_0 = 0.5$.
$\overline{d}$  is not significantly affected by the mixing conditions for a wide range of $C_0$.
On the other hand, $N_\mathrm{tot}$ is consistently higher for forced mixing than for unforced mixing. This can be explained by the difference in characteristic mixing velocity (either the mean drop speed $\bar{U}$ or the fluctuating velocity $\mathcal{U}$, see III.B) quantifying the mean number of droplets passing through the measurement plane and is an order of magnitude larger in the forced conditions than in unforced conditions. The dependence of $N_\mathrm{tot}$ on mixing is therefore due to enhanced flow in the forced mixing case, which causes more particles to pass through the measurement volume per unit time (and when $N_\mathrm{tot}$ is divided by $U$, forced and unforced conditions collapse on to the same curve; see Fig. \ref{fig:rescale_concentration}). 

To further analyze the various trends observed in Figure~\ref{fig:stats}, we now introduce a scaling analysis to understand the fraction of water vapor that is transformed into liquid and rationalize the size and number of droplets formed in each experiments.
First, we must relate $N_{\mathrm{tot}}$, the total number of droplets that pass through the PDA measurement region, to a absolute droplet concentration $C_{\mathrm{drop}}$ in units of particles/cm$^3$. We define $C_{\mathrm{drop}}$  as
\begin{equation}
    C_{\mathrm{drop}} = \frac{N_\mathrm{tot}}{A \, \mathcal{U}_{rms} \, \tau_f} 
    \label{eq:cdrop}
\end{equation}
where $A$ is the area of the measurement plane, $\mathcal{U}_{rms}$ is the characteristic speed of droplets passing through this measurement plane, 
and $\tau_f$ is the cloud lifetime. 
The area of the measurement region is known from the calibration of the PDA system, $A$ = 0.119 mm $\times$ 0.1196 mm, see \S~\ref{sec:methods:measurements:drops}. 
The definition of $C_{drop}$ therefore naturally encapsulated the importance of the initial saturation ratio and the mixing (flow) conditions through $\tau_f$ and $\mathcal{U}_{rms}$.

Prior to an expansion, the water vapor concentration is given by the initial saturation ratio, $S_0$, which determines the mass of liquid vapor available for condensation.
The pressure drop during the expansion, $\Delta p$, determines both changes in temperature and the partial pressure of water during the expansion, which combine to control changes in saturation ratio, leading to an estimated maximum $S_\textrm{max} $ typically above 3 or 4.
As such, it is reasonable to assume that all seeding aerosols will be activated and lead to droplet formation during an expansion, i.e.,\ $ C_\mathrm{drop} \approx C_0$.

Figure~\ref{fig:rescale_concentration} shows a plot of $C_{\mathrm{drop}}$ calculated from Eq.\ \ref{eq:cdrop} as a function of $C_0$ for all initial $S_0$ and mixing conditions. Error bars on measurements of $C_\textrm{drop}$ derive from uncertainties in $\tau_f$ and $\mathcal{U}_{rms}$. 
At intermediate concentrations ($4\times10^3$~\#/cm$^3$ $< C_0 < 8\times10^4$~\#/cm$^3$),  $C_\textrm{drop}$ falls close to the grey dashed line which has a slope of $C_\textrm{drop}/C_0=1$, suggesting that every solid aerosol in the chamber nucleates a droplet during the expansion.
The curves of $C_{\mathrm{drop}}$ deviate from the 1:1 line for low and high values of $C_0$.
Deviations at lower aerosol concentrations ($C_0 < 4\times10^3$ \#/cm$^3$) could be related the larger measurement error of PDA at lower particle counts, as well as related uncertainties in estimating the cloud lifetime (see Figure~\ref{fig:swarm_plot}). 
At high concentration ($C_0 > 8\times10^4$~\#/cm$^3$), the concentration of droplets observed undershoots the concentration of seeding aerosols, i.e. the fraction of activated aerosol particles is less than 1. This suggests that at these large $C_0$ values, rapid condensation onto numerous,  easily activated condensation nuclei  indeed  keeps the actual S in the chamber very close to 1, thus preventing the activation of other nuclei in the aerosol distribution.   Experimental uncertainties could also be partly responsible for $C_{\mathrm{drop}}< C_0$ : PDA is only sensitive to droplets larger than $1$~\textmu{m} so if smaller droplets are present, they would be missed by this analysis. And indeed, as detailed below when more droplets are formed they will be smaller on average, due to conservation of mass.

The mass of liquid water per unit volume that condenses following an expansion can be estimated as:
\begin{equation}
    \mathcal{L} = q_l \rho_f = \frac{\pi \overline{d}\,^3}{6} \, C_\mathrm{drop} \rho_w  
    \label{eq:water_content}
\end{equation} 
where $q_l$ is the specific liquid water content  (kg liquid water/kg humid air) of the air parcel after the expansion, $\rho_f$ is the density of the air parcel after the expansion, and $\rho_w$ is the density of liquid water.
$\rho_f$ can be found with the ideal gas law as $\rho_f= \pf/(\Rd\Tf)$. $q_l$ is given by $q_l=S_0 q^*(\To,\po)-q^*(\Tf,\pf)$ from conservation of total water (both liquid and vapor) and can be estimated using the expression for $q^*$ in Eq.\ \eqref{qvstar_romps}.
From Eq.\ \ref{eq:water_content}, we see that total mass of liquid water will scale as $C_\textrm{drop} \times\overline{d}^3$. This implies that when many droplets are formed they will tend to be smaller. 

From Eq.~\eqref{eq:water_content}, we also have $\overline{d}\propto \mathcal{L}^{1/3}$. Taking  $C_\mathrm{drop}~\approx~C_0$, we can rearrange Eq.\ \eqref{eq:water_content} to find that:
\begin{equation}
    \overline{d} =\left({\frac{6 q_l \rho_f}{\pi C_\textrm{drop}\rho_w}}\right)^{1/3} \approx \left({\frac{6 q_l \rho_f}{\pi C_0\rho_w}}\right)^{1/3}
    \label{eq:size_vs_c0}
\end{equation}
The three red dashed lines in Figure~\ref{fig:stats}~(b) represent least-square linear regression fits of the function $\overline{d} = \mathcal{F}C_0^{-1/3}$ to our experimental data, with good agreement.

We can qualitatively compare the droplet size scaling with $C_0$ to the work of \citet{ray_shaw2023fast} 
in the context of the $\Pi$ chamber experiments. The $\Pi$ chamber operates in statistically steady-state conditions, with a constant aerosol injection rate of $\dot{n}_\mathrm{inj}$, compensated by loss at the walls and due to sedimentation. The mean size $\overline{d}$ is then related to the rate of droplet nucleation assumed proportional to the injection rate, $\overline{d}$ $\propto \dot{n}_{\mathrm{inj}}^{-1/3}$ \cite{ray_shaw2023fast}. 
In our cloud chamber, the expansion is a transient phenomenon. But if we assume $ \dot{n}_\mathrm{inj} \propto C_0$, our scaling argument for the droplet size with aerosol concentration given in Eq.\ \eqref{eq:size_vs_c0} and our data (Figure \ref{fig:stats}) agrees with the steady state results of \citet{ray_shaw2023fast}.
The agreement provide confidence that the ability to rapidly probe a wide range of of experimental conditions in our experimental apparatus can provide physical insight into understanding cloud microphysics.



\subsection{Holographic measurements of nucleation onset and time-resolved droplet growth} \label{subsec:drop_diam}

\begin{figure}
    \centering    
    \includegraphics[width=\linewidth]{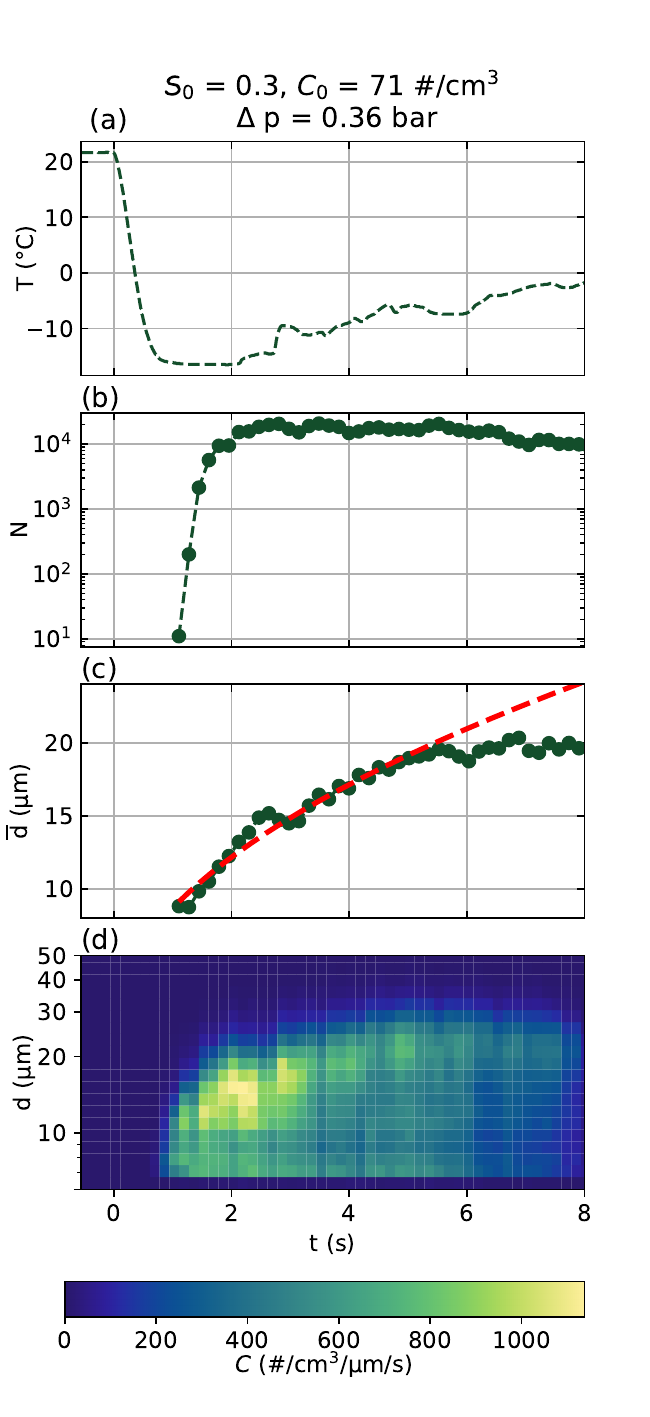}
    \caption{Time evolution of (a) temperature, (b) droplet number, (c) mean droplet diameter, and (d) droplet size distribution during an expansion with $S_0 = 0.3$, $C_0 = 71$~\#/cm$^3$ and $\Delta p = 0.36$~bar. The expansion is triggered at $t = 0$~s. The thermodynamic conditions in the chamber are close to the onset of droplet nucleation. Droplet numbers and diameters are recorded using an inline holographic system recording at 600~Hz. The time bins in subplots (b) and (c) are spaced so $\Delta t = 0.17$~s. The red line in subplot (c) shows a fit to the droplet diameter data from $t = 1.2$ to 5.5~s of the function $d(t)=\sqrt{d_0^2 + a(t-t_0)}$ where $d_0 = 6$~\textmu{m}, $t_0 = 0.5$~s, and $a = 73$~m$^2$/s. 
    The color map in subplot (d) shows the concentration of droplets in each bin, $C$, per unit volume, per micron, per second.}
    \label{fig:dist_evolution}
\end{figure}

Finally, we demonstrate the capabilities of the holographic system in tracking droplet nucleation and growth. The holographic system has the advantage of being able to probe a much larger volume than the PDA and average over the chamber length, therefore accessing more dilute regimes, closer to nucleation threshold and closer to the droplet concentration observed in actual clouds. It is important to note that thanks to the large sampling volume (8.44~cm$^3$), the holographic system measures more individual droplets and achieve better resolved statistics, allowing to track the drop size distribution in time.

We report the temporal evolution of droplet size distribution tracked with holography in a single expansion in Figure~\ref{fig:dist_evolution}, with panel b showing the number of drops detected, c the mean droplet diameter, and d the droplet size distribution.
The experiment is performed with an initial saturation ratio of $S_0 = 0.3$, an initial aerosol concentration of $C_0 = 71$ \#/cm$^3$, and a pressure drop of $\Delta p = 0.36$ bar under unforced mixing conditions. The minimum temperature is then close to -16 degrees (fig. 12a).
This combination of $\Delta p$ and $S_0$ is chosen to be relatively close to the threshold of droplet nucleation, while we  work with a low concentration of seeding aerosols to avoid excessively high droplet concentrations that can obscure the holographic data. 

The first droplets are detected by the holographic system approximately 1 second after the temperature drop.
In a first regime, occurring 1 to 2.5 seconds after expansion, we observe a rapid increase in droplet number detected in the measurement volume up to $10^4$ (fig. 12b), together with an increase in the diameters, from a few microns to 20 microns (fig. 12c and d). The droplet diameters appear to grow as $\sqrt{t}$, consistent with diffusive growth. We note that droplet growth beyond 20-25 microns would not occur by diffusive growth in the atmosphere but through collision processes.
After 2-3 seconds, the number of drops plateaus before slowly decreasing, while the size keeps increasing until about 6~s. 
The concentration of liquid droplets is calculated by dividing the number of measured droplets by the holographic system measurement volume (8.44~cm$^3$). The maximum droplet concentration is approximately 22~$\pm$~2.1\#/cm$^3$ and is reached at $t = 2.0$~s, during the first droplet growth regime. The value of the maximum concentration is of the same order of magnitude than the seeding concentration of aerosols. 

These findings highlight our ability to directly measure the growth of droplets during an expansion using holography in low concentration regimes. We note that the absolute number of drops detected here is much higher than when using the PDA in the previous section because of the much larger detection volume; while the actual concentration is actually smaller. Overall, the combined use of the PDA and the holography allow to probe droplet nucleation and growth over a wide range of aerosol concentration.

\section{Conclusions \label{sec:conc}}
We have presented a new experimental facility designed to study processes relevant to aerosols and cloud microphysics. The experimental principle of our facility is that of an expansion chamber, but with the added ability to induce various turbulent flows. 
The experimental facility comprises two vacuum-sealed chambers which are capable of reproducing an adiabatic-like expansion, where the gas from the aerosol-nucleation chamber is allowed to expand into evacuated expansion chamber, causing a sudden increase in saturation ratio.
We demonstrate precise control over the initial conditions of the expansion, including the initial saturation ratio, $S_0$, aerosol concentration, $C_0$, and air mixing conditions.

Our specific experimental design allows us to achieve a high repetition rate of experiments to probe a wide range of conditions, and ample optical access to perform high frequency real-time measurements. This report represents an initial demonstration of these capabilities. First, we characterize the chamber thermodynamics, finding the minimum temperature during the expansion in good agreement with theoretical predictions for both dry and moist conditions. We then present a discussion of droplet nucleation and growth in the presence of aerosol particles for a range of experimental parameters. We characterize the total droplet number and mean droplet diameter throughout expansions. We observe the classic competition between number of droplets and droplet size for a given amount of liquid vapor: for more aerosol seeds, more droplets are observed and they are on average smaller. The droplet concentration is shown to scale linearly with the initial aerosol concentration, while the mean droplet size decreases with seeding aerosol concentration. Simple scaling laws based on thermodynamical principle can describe these trends. Finally, we demonstrate the ability to characterize droplet growth at conditions near the onset of nucleation using inline holography.

The key advantage of our facility is our high experimental repetition rate and our use of high-speed measurement techniques that enable us to conduct systematic parametric studies more efficiently than larger facilities.
Future studies will involve detailed investigations of the interplay of homogeneous and heterogeneous droplet nucleation for various initial conditions and seeding aerosols. 
We are also implementing high-speed infrared absorption  measurements to directly measure the changing water vapor concentration (and therefore saturation ratio) throughout each expansion, coincident with holographic measurements. 
We will also consider the role of turbulent mixing of different intensities and anisotropy. Finally, the chamber walls are equipped with a cooling system which will permit us to explore lower-temperature conditions and the nucleation and growth of ice crystals in future work.

\section{Acknowledgments}
This work was supported by a seed grant from the Simons Foundation (Award \#1019423), and Princeton University internal funds from the Princeton Catalysis Initiative and the High Meadows Environmental Institute's Climate and Energy Grand Challenge Award granted to L.D. and M.L.W. We thank Clare Singer from NOAA-GFDL for providing comments and performing the internal review. 


\bibliography{main_text}

\end{document}